\documentstyle[11pt,newpasp,twoside,epsf]{article}
\def\tauad{\tau_{\rm ad}}
\def\BP{Ba\-lles\-te\-ros-Paredes}
\def\ls{\lambda_{\rm s}}
\def\tauff{\tau_{\rm ff}}
\def\VS{V\'azquez-Semadeni}

\def\edcomment#1{\iffalse\marginpar{\raggedright\sl#1\/}\else\relax\fi}
\reversemarginpar

\begin{document}

\title{The Turbulent Star Formation Model. Outline and Tests}

\author{Enrique V\'azquez-Semadeni}
\affil{Centro de Radioastronom\'\i a y Astrof\'\i sica, UNAM, Campus
Morelia, Apdo. Postal 3-72, Morelia, Michoac\'an, 58089, M\'exico}

\begin{abstract}
We summarize the current status of the turbulent model of star
formation in turbulent molecular clouds. In this model, clouds, clumps
and cores form a hierarchy of nested density fluctuations caused by
the turbulence, and either collapse or re-expand. Cores that collapse
can be either internally sub- or super-sonic. The former cannot
further fragment, and can possibly be associated with the formation of a
single or a few stars. The latter, instead, can undergo turbulent
fragmentation during their collapse, and probably give rise to a
cluster of bound objects. The star formation efficiency is low because 
only a small fraction of the density fluctuations proceed to
collapse. Those that do not may constitute a class of ``failed'' cores 
that can be associated with the observed starless cores. ``Synthetic''
observations of cores in numerical simulations of non-magnetic
turbulence show that a large fraction of them have subsonic internal
velocity dispersions, can be fitted by Bonnor-Ebert column density
profiles, and exhibit ``coherence'' (an apparent independence of
linewidth with column density near the projected core centers), in
agreement with observed properties of molecular cloud cores.
\end{abstract}

\section{Introduction} \label{sec:intro}

Two key questions related to star formation are a) What is the
origin and nature of star-forming (and non-star-forming, if they
exist) cores in turbulent molecular clouds? b) The star formation
efficiency (SFE), measured as the fraction of a molecular cloud's mass 
that ends up in stars during its lifetime, is low, of order 5-10\%
(see, e.g., the review by Evans 1999). Why? Different answers to these
questions are given in the main two competing models of star
formation. In the so-called ``standard'' model (Mouschovias 1976; Shu
1977; see the review by Shu, Adams \& Lizano 1987), low-mass
star-forming cores begin their
lives as larger, magnetostatic, magnetically subcritical,
partially-ionized clumps of low density 
contrast relative to their parent cloud. These slowly lose magnetic
support and contract as the neutrals ``slip'' through the ions in a process
commonly referred to as ``ambipolar diffusion'', with characteristic
time scale $\tauad$, until they finally become magnetically
supercritical and proceed to gravitational collapse. The
low SFE is explained in this case because $\tauad$
is much larger than the free-fall time scale $\tauff$, and thus constitutes
the effective time scale for star formation,
rather than $\tauff$. High-mass star
formation is assumed to occur through the collapse of clumps that
agglomerate enough mass to become supercritical and thus collapse on a
time scale $\tauff$, although clumps this massive are assumed to be
rare. However, this model suffers from a number of   
shortcomings (see, e.g., Mac Low \& Klessen 2003; Hartmann, this
conference). In particular, it does not address  
the formation of the clumps, which are simply taken as intial
conditions, nor how can such
magnetostatic structures survive in the turbulent
environment of their parent clouds. 

In the competing, more recent turbulent scenario (e.g., Elmegreen 1993;
Padoan 1995; \BP, \VS\ \& Scalo 1999a, \BP, Hartmann \& \VS\ 1999b;
Klessen, Heitsch \& Mac Low 2000; Heitsch, Mac Low \& Klessen 2001;
Hartmann, \BP \& Bergin 2001; Padoan \& 
Nordlund 2002; \VS, \BP\ \& Klessen 2003; see also the reviews by \VS\ 
et al.\ 2000; Mac Low \& Klessen 2003) the answer to question (a) above 
is that the cores
are turbulent density fluctuations in the molecular clouds (which have 
typical rms Mach numbers $M_{\rm s} \sim 10$) and that their
statistical properties are therefore determined by the turbulent
parameters in the parent cloud. The clouds are globally supported by
turbulence, and in fact may be transient with short lifetimes (\BP\ et
al.\ 1999b; Hartmann et al.\ 2001), but locally the density
fluctuations induced by the turbulence may or may not collapse
depending on their particular energy balance (\BP\ \& \VS\ 1997; \VS,
Shadmehri \& \BP\ 2002). Then, the answer to 
question (b) above in the turbulent model is that the SFE is low
because the cores contain only a fraction of the total cloud mass,
only a fraction of them undergoes collapse, and, as is also the case
in the standard model, only a fraction of the cores' mass is involved
in the collapse. That is, only a small fraction of the
globally-turbulence-supported cloud is able to undergo collapse,
through a kind of ``turbulent colander''.
Note that, while magnetic support is a key
concept in the standard model of star formation, in the turbulent
model it is not indispensible, as delay of the collapse is not necessary.


In this paper we first summarize the current status of
the turbulent model of star formation, together with key work that has
led to its formulation (\S 2), and then we discuss evidence that
indeed turbulence 
can produce structures which reproduce some important observational
properties of ``quiescent'' molecular cloud cores (\S 3), thus
implying a high plausibility for the turbulent model. In this
paper, we assume that the turbulence in molecular clouds is
continuously driven, as otherwise it would decay in roughly one
crossing time (Mac Low et al.\ 1998; Stone et
al.\ 1998; Padoan \& Nordlund 1999). This is in fact expected if molecular
clouds are an intermediate-scale 
part of the turbulent cascade in the ISM, since energy is then
continuously cascading {\it through them} from the large 
energy injection scales to the small dissipative ones.

\section{The turbulent model of core and star formation}
\label{sec:turb_scen} 

It was already forseen by von Weizs\"acker (1951) and Sasao (1973)
that compressible turbulence in gas clouds must have a dual role:
turbulent modes of size scale $\lambda$ provide ``pressure'' towards
structures with sizes $>  \lambda$, while simultaneously they induce the
formation of large-amplitude density fluctuations of sizes $<
\lambda$. The latter 
process is generally called ``turbulent fragmentation''. Note that, in
order to be highly compressible, the turbulence must be supersonic, as
is observed (see, e.g., Zuckerman \& Palmer 1974), and therefore
it overwhelms the thermal support at all but the smallest 
scales. This implies that turbulence-supported clouds may have masses
much larger than their thermal Jeans mass.

Sasao (1973) already pointed out that the compressed regions may be driven to
collapse, and Hunter \& Fleck (1982) showed that within them the local
Jeans length is strongly reduced. This implies that local collapsing
cores generally may have masses smaller than even the thermal Jeans
mass at the mean cloud density, and so their masses are in general
much smaller than their parent cloud's mass. Note additionally that the
free fall time in the cores is given by $\tauff \sim
(G \rho_{\rm core})^{-1/2}$, and is therefore shorter than that of
their parent cloud by a factor 
$\sqrt{\rho_{\rm cloud}/\rho_{\rm core}}$. This is the opposite to the
case of linear (small-amplitude) perturbations, in which the fastest
collapsing scale is the largest one (Larson 1985).


Moreover, not all the density fluctuations necessarily proceed to
collapse; whether they do or do not depends on whether they acquire
enough mass that their gravitational energy overwhelms all other forms
of energy that provide support. 
If it does not, the fluctuations re-expand, as first shown numerically
by \VS, Passot \&
Pouquet (1996) using non-isothermal, non-magnetic simulations
subject to 
turbulent driving. In these, many generations of density fluctuations (cores)
were seen to appear and disappear, until finally a (rather improbable) 
strong enough compressive 
event managed to produce a gravitationally unstable core, which
then proceeded to collapse in roughly a {\it local} free-fall time.
Moreover, the global density maximum in the simulations was seen to
fluctuate chaotically in time (showing also the
non-formation of any hydrostatic cores), until it suddenly began to
increase at an accelerated pace, corresponding to the local, almost
instantaneous collapse of a ``core''. 


Padoan (1995; see also Padoan \& Nordlund 2002; \VS\ et al.\ 2003)
proposed that the only fragments that collapse are those which
are far enough 
down the turbulent cascade that their internal velocity dispersion is
subsonic,\footnote{Note that the existence of a subsonic range is
simply the consequence of a turbulent cascade, and does not
necessarily imply that 
subsonic cores are at the scale of turbulence dissipation.} and that in
addition are gravitationally unstable. This scenario has the
implication that, if one 
considers a succession of turbulent regimes in which subsonic velocity
dispersions occur at progressively smaller scales, the fraction of the
mass available for collapse, and therefore the SFE, should also be
progressively smaller, because smaller regions contain smaller
fractions of the total mass. 

This suggestion was tested numerically by \VS\
et al.\ (2003). For a series of SPH simulations of
turbulent clouds, they empirically measured the ``sonic scale'' $\ls$, i.e.,
the scale at which the turbulent velocity dispersion equals the sound
speed. Then they also operationally defined and measured the SFE as
the fraction of mass in collapsed sink particles after 
two turbulent crossing times. With these data, they showed that the
SFE correlates well with $\ls$, and in fact scales as SFE $\approx
\exp(-\lambda_0/\ls)$, where $\lambda_0$ is a suitable reference
scale.
This suggests that indeed the sonic scale is one fundamental 
parameter in determining the SFE. Additionally, \VS\ et al.\ (2003)
also showed that the SFE does {\it not} correlate well with the
driving scale,
and that previous results suggesting so (Klessen et 
al.\ 2000) were only an artifact of maintaining the same total kinetic 
energy in runs with different driving scales, as this caused higher
energies per unit wavenumber interval in runs forced at smaller
scales, decreasing the sonic scale.


However, it is also clear that not only subsonic cores can collapse
(if they are super-Jeans), but also supersonic cores, if gravity anyway
overwhelms all other forces. In this case, not contemplated by either
Padoan (1995), Padoan \& Nordlund (2002) nor \VS\ et al.\ (2003),
turbulence will be able to still produce fragments, but not support
the cloud as a whole. Given the shorter free-fall time of the smaller,
nonlinear scales, these will collapse first. As soon as they do,
they will collectively become a non-dissipative 
system, and thus will not collapse to a singularity, but instead form a 
bound cluster of collapsed objects (see Bate, this volume). Thus, one
can tentatively 
associate the global collapse of an unsupported supersonic region with 
the ``clustered'' mode of star formation, while the individual
collapse of subsonic regions may be associated with the ``isolated''
mode, in which each core may produce one or a few stars. 

Finally, there is the issue of the cores that do not collapse, but
instead rebound after the compression is over, and merge back with
their surronding medium. \VS\ et al.\ (2002) have performed a simple
calculation to estimate the re-expansion time
$\tau_{\rm exp}$, defined as the time it takes to
double the initial radius. This is shown in fig.\ 1 as a function of
the core radius at maximum compression, with
$\tau_{\rm exp}$ given in units of the free-fall time, and the radius
in units of the equilibrium radius at which internal pressure balances
self-gravity. We see that the minimum re-expansion time is of the
order of a few free-fall times, or $\sim 1$ Myr. This picture is
consistent with observational estimates of core lifetimes, and, in
particular, the proposed existence of failed cores is consistent with
the known existence of a large number of starless cores (at least half
of the total; e.g., Lee \& Myers  1999). These cores have traditionally been
considered to be in a pre-protostellar evolutionary stage, but in the
turbulent scenario they may well never form stars (Taylor, Morata \&
Williams 1996; \VS\ et al.\ 2002).

\begin{figure}
\plotfiddle{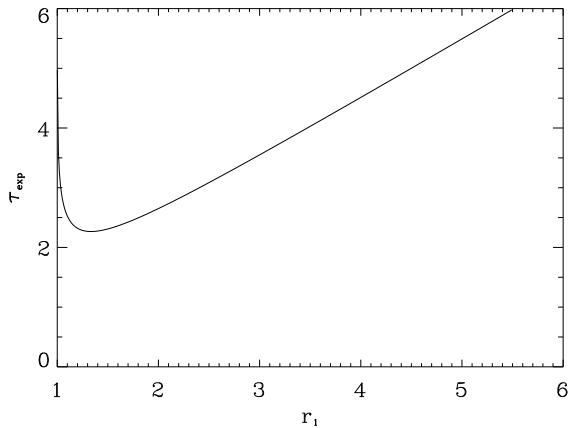}{2.5in}{0}{50}{50}{-165}{-175}
\caption{Re-expansion time of a core, in units of the free-fall time,
defined as the time necessary to
double the initial core's radius, as a function of the initial radius
$r_1$, normalized to the equilibrium radius.}
\label{fig:re_exp}
\end{figure}


We conclude this section by noting that it has all been based on the
concept and numerical simulations of the {\it non-magnetic}
case. Thus, within the context of the turbulent star
formation scenario, the magnetic field is not an indispensible
ingredient. The isolated and clustered modes of star formation, as
well as the low SFE, can be well understood in terms of purely hydrodynamic,
self-gravitating turbulence. We expect the presence of a magnetic
field to introduce only quantitative modifications to this scenario.

\section{Comparisons with observations} \label{sec:tests}

In order to assess the feasibility of the turbulent model of star
formation, it is necessary to compare its results with the
observations. 
A question that is frequently
asked in relation to the turbulent model is whether it can explain the 
relatively quiescent nature of observed cores, where by ``quiescent'' it is
normally meant subsonic, or transonic at most. However,
as discussed in \S 2, the
existence of subsonic velocity dispersions at small enough scales is a 
natural consequence of the turbulent cascade, and in fact, the ``sonic 
scale'' appears to play
a key role in the control of the SFE, as described there. 

Additionally, Barranco \& Goodman (1998) have observationally found
that the linewidth in a (small) sample of cores exhibits the
interesting property of becoming independent of the offset from the
core center at small offsets. This asymptotic value is slightly larger 
than the thermal value, and was interpreted by those authors as an
indication that the {\it non-thermal} velocity dispersion becomes
scale-independent in 
the cores, saturating at a slightly subsonic level. They referred to
this phenomenon as ``coherence'' of the 
velocity field in the innermost regions of the cores, within scales
$\la 0.1$ pc. 

Finally, several recent observational works have 
fitted, with various degrees of success,
Bonnor-Ebert (BE; Ebert 1955; Bonnor 1956) column density profiles to
observed cores. BE spheres are hydrostatic, non-magnetic
configurations, bounded by a hot, tenuous medium that contributes
external pressure, but not to the gravitational potential of the
system. The presence of such a hot confining medium is indispensible
for the stability of the system, since otherwise the configurations
need to smoothly extend to infinity, and in this case are generally
unstable, as is, for example, the case of the singular isothermal sphere.
\VS\ et al.\ (2002) have pointed out that such configurations
cannot be realized in single-temperature media (as molecular clouds
appear to be), since in this case the requisite of the hot, tenuous
confining medium cannot be fulfilled. However, it is then puzzling
that the observations seem to reasonably fit the column density
profiles of BE spheres reasonably well.

In a couple of recent papers, we have addressed these two questions.
\BP\ et
al.\ (2003) considered a sample of 378 projections of the cores in the 
numerical simulations onto a hypothetical plane of the sky (POS). Column
density profiles, 
averaged over the position angle on the POS, were produced, and the
scatter around this average was recorded. Then, BE column density
profiles were fitted to the average profiles, and accepted if the
error between the fitted and average profile was smaller than the
scatter observed in the profiles contributing to the average.
Otherwise, the fits were rejected. This procedure yielded 65\% of
``acceptable'' BE fits to the profiles, showing that the transient
cores in the simulations have nevertheless angle-averaged profiles
that often resemble BE ones. This can probably be attributed to the
strong smoothing introduced by the angle-averaging.

The issue of a sub- or transonic nature of the cores has been
addressed by Klessen et al.\ (2003), who investigated the velocity
structure for the core sample studied by \BP\ et al.\ (2003). 
They found that, in simulations 
with gravity turned off, roughly half the cores have subsonic velocity
dispersions. In simulations with gravity on, the number of
subsonic cores decreases, but still the vast majority has velocity
dispersions that do not exceed twice the sound speed, and roughly
10\% have subsonic velocity dispersions. Moreover, the
velocity dispersion was measured in this plot out to column densities
of 1/5 times the maximum, which probably increases the measured
velocity dispersion, in comparison to the standard procedure of
measuring out to the FWHM only. 
Thus, the simulations are entirely
consistent with the fact that molecular cloud cores are generally
transonic and often even subsonic.


Klessen et al.\ (2003) also probed the cores in the simulations for
``coherence''. Figure 2 shows the velocity dispersion
vs.\ column density along all lines of sight through a few randomly
selected cores, clearly 
showing a trend to acquire a constant value, often subsonic, towards
high column densities.  Thus, the cores in the simulations also
exhibit ``coherence'' when projected on the POS. 
Since such independence of scale of the turbulent motions has no
counterpart in normal turbulent flows, Klessen et al.\ (2003)
suggested that the ``coherence'' may be an observational effect,
probably due to the fact that near the projected core centers, the
linewidth along a given line of sight is dominated by its length
rather than by its typical scale on the POS (i.e., the contributing
material lies within a cylinder that is longer than it is wide). If
the column length does not vary much around the projected core
maximum, neither will the linewidth, causing the apparent
``coherence''. 

\begin{figure}
\plotone{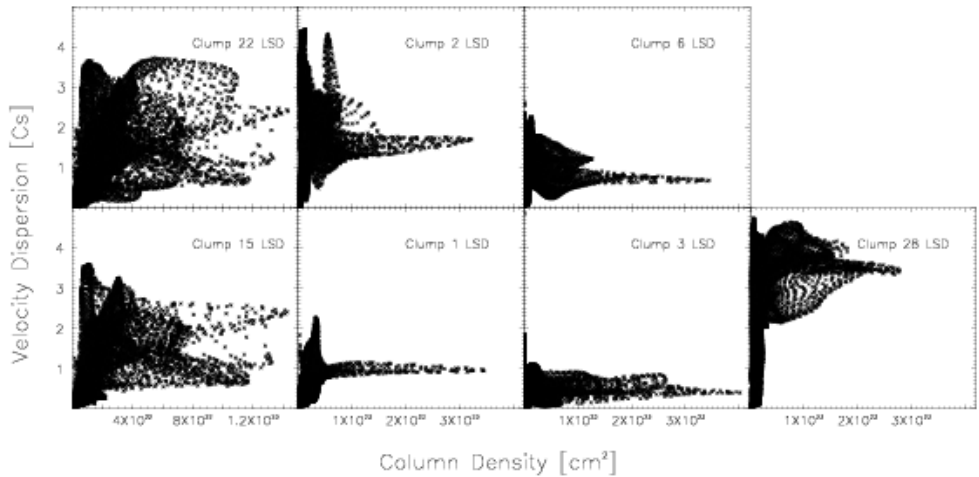}
\caption{Velocity dispersion
vs.\ column density along all lines of sight through a few randomly
selected cores in one turbulent simulation driven at large
scales. Compare to fig.\ 3 of Barranco \& Goodman (1998).}
\label{fig:coherence}
\end{figure}

\section{Conclusions} \label{sec:conclusions}

In this paper we have given a global outline of the turbulent model of 
molecular clouds and star formation, and then summarized a number of
tests of the simulations showing that the cores in the non-magnetic
simulations studied exhibit similar density and velocity features 
to those observed in real cores, strongly supporting the validity of
the turbulent model of molecular clouds. Provided that the turbulence in
molecular clouds is continuously driven, the model explains in a
natural way the origin and structure of the cores as transient
turbulent density 
fluctuations in the clouds, and the low efficiency of star formation,
since a) the cores contain a small fraction of the cloud's mass, and
b) not all cores are destined to collapse. Those that do not, rebound
and merge back into their environment, and may generally correspond to
the observed ``starless'' cores. This scenario does not
necessarily rely on the magnetic field to operate, although the
presence of such a field will most certainly introduce quantitative
corrections into the model. Although further testing is still
clearly necessary, and the role of the magnetic field needs to be
quantified, a quantitative theory of star formation in 
turbulent molecular clouds appears to be under way.

\acknowledgements
The author gratefully acknowledges useful comments and criticisms from
Susana Lizano and Javier Ballesteros-Paredes.
This work has received financial support from CONACYT grant 36571-E.

\end{document}